# Detergency and its implications for oil emulsion sieving and separation

*Thomas M. Schutzius,[a,†] Christopher Walker,[a,†] Tanmoy Maitra,[a] Romy Schönherr,[a] Christos Stamatopoulos,[a] Stefan Jung,[a] Carlo Antonini,[a] Hadi Eghlidi,[a] Julie L. Fife,[b] Alessandra Patera,[b] Dominique Derome,[c] Dimos Poulikakos [a,\*]*

[a] Laboratory of Thermodynamics in Emerging Technologies, Department of Mechanical and Process Engineering, ETH Zurich, Sonneggstrasse 3, CH-8092 Zurich, Switzerland

[b] Swiss Light Source, Paul Scherrer Institut, CH-5232 Villigen PSI, Switzerland

[c] Laboratory for Multiscale Studies in Building Physics, Swiss Federal Laboratories for Materials Science and Technology, Empa, Überlandstrasse 129, CH-8600 Dübendorf, Switzerland

**Keywords:** Oil-water separation, Emulsion separation, Detergency, Surfactants, Detergents, Filtration, Meshes, Synchroton X-ray microtography

[†] Denotes equal contribution

[*] To whom correspondence should be addressed.

Prof. Dimos Poulikakos
ETH Zurich
Laboratory of Thermodynamics in Emerging Technologies
Sonneggstrasse 3, ML J 36
CH-8092 Zürich
SWITZERLAND
Phone: +41 44 632 27 38
Fax: +41 44 632 11 76
dpoulikakos@ethz.ch

**Abstract**

Separating petroleum hydrocarbons from water is an important problem to address in order to mitigate the disastrous effects of hydrocarbons on aquatic ecosystems. A rational approach to address the problem of marine oil water separation is to disperse the oil with the aid of surfactants in order to minimize the formation of large slicks at the water surface and to maximize the oil-water interfacial area, which governs the biodegradation rate. Here we investigate the fundamental wetting and transport behavior of such surfactant-stabilized droplets and the flow conditions necessary to perform sieving and separation of these stabilized emulsions. We show that, for water soluble surfactants, such droplets are completely repelled by a range of materials (intrinsically underwater superoleophobic) due to the detergency effect; therefore, there is no need for surface micro/nanotexturing or chemical treatment to repel the oil and prevent fouling of the filter. We then simulate and experimentally investigate the effect of emulsion flow rate on the transport and impact behavior of such droplets on rigid meshes to identify the minimum pore opening ($w$) necessary to filter a droplet with a given diameter ($d$) in order to minimize the pressure drop across the mesh—and therefore maximize the filtering efficiency, which is strongly dependent on $w$. We define a range of flow conditions and droplet sizes where minimum droplet deformation is to be expected and therefore find that the condition of $w \approx d$ is sufficient for efficient separation. With this new understanding, we demonstrate the use of a commercially available filter—without any additional surface engineering or functionalization—to separate oil droplets ($d < 100$ µm) from a surfactant stabilized emulsion with a flux of ~11,000 L m$^{-2}$ hr$^{-1}$ bar$^{-1}$. We believe these findings can inform the design of future oil separation materials.

**Introduction**

       Petroleum hydrocarbon releases into marine ecosystems from industrial accidents or oil tanker sinking are disastrous for aquatic ecosystems [1, 2]. Strategies to mitigate the adverse effects of such events vary depending on circumstances (e.g. biodegradation, skimming, controlled burns, siphoning, etc.), but the goal is the same—complete removal of hydrocarbons. Reduction of oil substances dispersed in water is a challenging problem, in part due to the number of different ways that the system can manifest itself—separate phase, two-phase emulsion, or solubilized [3]. In the Deep Water Horizon spill, one of the strategies was to inject dispersant (COREXIT 9500) at the wellhead that was 1500 m below water in order to disperse the oil at-depth and prevent large slicks [1]. Such an approach naturally produces a two-phase emulsion. While smaller oil droplets are neutrally buoyant, larger ones will eventually migrate to the surface of the water and, if collected, will need to be separated.

      Previous work on oil-water emulsion separation has emphasized the role of filters for promoting coalescence [4], while more recent work, mostly grounded in wettability engineering research, has emphasized surface micro/nanoengineering to impart filters with underwater superoleophobicity to prevent fouling in filtration ranging from micro- to ultrafiltration [5, 6]. While such approaches are logical for simple oil-water mixtures, it is not obvious that such surface engineering is necessary for surfactant stabilized oil-in-water emulsions. Taking inspiration from a classic experiment done in the 1930s by Adam [7], whereby the so-called "rolling up" mechanism was identified [8], we investigate this phenomenon and test its implications for separation in microfiltration, whose findings, we believe, will help govern nano- and ultrafiltration design. The "rolling-up" mechanism, Figure 1, shows the wetting behavior of hexadecane to a standard fluoropolymer (poly(vinylidene fluoride)) without and with the

presence of a surfactant, namely sodium dodecyl sulfate (SDS). In the former case, the receding contact angle ($\theta_r^*$) is $\approx 0°$—practically superoleophilic—while in the latter case it is $>150°$, practically superoleophobic. It should be noted that this extreme non-wetting state is achieved in the absence of any surface micro/nanotexturing, which is usually required to observe such behavior [9]. While oil-solid interactions in micellar solutions are well-documented [7, 10-14], such knowledge has yet to be applied to the design of meshes for separation of surfactant-stabilized oil-water emulsions [6, 9, 15-17]. Therefore, while previous work has been done on the effect of pore size on pressure drop and separation [4, 18], the focus of the present investigation is to understand this surfactant-induced anti-wetting mechanism, which is referred to here as detergency—the process whereby soils, liquid or solid, are removed from a substrate material [10]—and to see its implications for oil separation from surfactant stabilized emulsions by studying droplet-mesh interactions at the single droplet level.

**Results and Discussion**

*Surfactant Influence on Wetting Behavior: The Role of Detergency*

To better understand the effect of substrate material on surfactant-induced underwater oleophobicity, we began by observing the influence of an anionic surfactant, sodium dodecyl sulfate (SDS), on the wetting behavior of an oil (hexadecane) droplet on a variety of substrates in an aqueous medium. We conducted two sets of contact angle measurements in pure water and in a micellar solution sufficiently above the critical micelle concentration (0.03 mol L$^{-1}$ in water). Each set consisted of both receding ($\theta_r^*$) and advancing ($\theta_a^*$) contact angle measurements on a variety of clean, smooth surfaces made of or coated with different materials.

Figure 2a shows a plot of $\theta_a^*$ and $\theta_r^*$ for hexadecane underwater vs. surface composition and it is clear that these values depend greatly on the surface type [19]. For the sake of discussion, we will concentrate on the two tested surfaces that have the most oleophobic and oleophilic wetting property, which are borosilicate glass (further referred to as glass) and poly(vinylidene fluoride) (PVDF), respectively. Due to the chemical nature of oxygen-plasma treated glass, its surface has a strong affinity for water [20]; therefore, it is expected to exhibit underwater oleophobicity. PVDF, on the other hand, is hydrophobic [21], so here one would expect that, in a water environment, PVDF would exhibit oleophilic behavior, as shown by Figure 2a. All of the other surfaces tested here exhibit $\theta_a^*$ and $\theta_r^*$ values that fall in-between those measured for PVDF and glass, with their values of $\theta_r^*$ being low enough to consider them oleophilic (< 90°). The high contact angle hysteresis can be explained by the fact that the surfaces all have some degree of roughness (see **Figure S1** for micrographs of the surfaces).

Figure 2b shows the effect of the addition of surfactant (0.03 mol L$^{-1}$) to the aqueous environment on $\theta_a^*$ and $\theta_r^*$ for the same surfaces studied above. In all cases, after the addition of surfactant, the surfaces became underwater superoleophobic. To understand the wetting behavior of hexadecane on glass and PVDF, where glass stayed superoleophobic and PVDF underwent an oleophilic to superoleophobic wetting transition, it is instructive to refer to the Young–Dupré equation. Here, the equation is applied to an oil droplet in a water environment in contact with a solid surface

$$\cos\theta = \frac{\gamma_{sw} - \gamma_{so}}{\gamma_{ow}}, \tag{1}$$

where $\theta$ is the intrinsic contact angle of an oil droplet on the surface and $\gamma_{sw}$, $\gamma_{so}$, and $\gamma_{ow}$ are the solid-water, solid-oil, and oil-water interfacial energies, respectively. The nature of surfactants is to positively adsorb at the interface between polar and non-polar materials causing the associated $\gamma$ to decrease. Based upon the nature of the interfaces, we can reason as to whether or not the surfactant molecules positively adsorb there and ultimately whether or not the respective $\gamma$ should be affected [7]. For the present system (oil-water-solid), a reduction in $\gamma_{ow}$ is expected due to the nonpolar and polar nature of oil and water, respectively; here, SDS migrates from the bulk water phase to the interface. Experimentally, we observed a significant reduction of $\gamma_{ow}$ with the addition of surfactant. (See Figure S2 for the effect of surfactant type (SDS and Triton-X 100) and concentration (between 0 and 0.04 mol L$^{-1}$) on $\gamma_{ow}$ for two oils, hexadecane and FC-770, which are less and more dense than water, respectively.) For example, when going from 0 to 0.03 mol L$^{-1}$ SDS in water, $\gamma_{ow}$ for a water-hexadecane interface reduced by ~60%. To understand the impact of the presence of surfactant on $\gamma_{sw}$ and $\gamma_{so}$, it is instructive to refer to the Gibbs adsorption equation, which relates the change in surface energy with the chemical potential of a single surfactant component and a surfactant adsorption density, $\Gamma$, combined with Eq. (1) to yield [22, 23]

$$\frac{d(\gamma_{so} - \gamma_{sw})}{d\gamma_{ow}} = \frac{\Gamma_{so} - \Gamma_{sw}}{\Gamma_{ow}}. \qquad (2)$$

In the case of a glass surface, previous work for a system with a high-surface-energy solid, an oil, and an anionic surfactant has shown that $\Gamma_{so} > \Gamma_{sw}$ [23], therefore, with the addition of surfactant, $\gamma_{so}$ is expected to decrease relative to $\gamma_{sw}$. From Eq. (1) we see that, in order to maintain underwater superoleophobicity, the right-hand-side of the equation should remain constant, which is possible if $\gamma_{sw}$ is sufficiently small relative to $\gamma_{so}$ prior to the addition of surfactant so that it remains smaller after the addition ($\Delta\gamma < 0$). Also required is that $\gamma_{ow}$ experiences a proportional decrease compared with $\gamma_{so}$. Due to the polar nature of both glass and water, it is conceivable that $\gamma_{sw}$ is relatively small, even with respect to $\gamma_{so}$ after surfactant adsorption. As for the simultaneous decrease of $\gamma_{ow}$ and $\gamma_{so}$ with surfactant adsorption, we expect $\gamma_{so}$ to decrease ($\Gamma_{so} > \Gamma_{sw}$) and from experiments we know that $\gamma_{ow}$ for a water-hexadecane interface is reduced (see Figure S2). Together, this is our interpretation of how a polar material (glass) can maintain underwater superoleophobicity after exposure to a surfactant.

The PVDF substrate represents a different case. In Figure 2b we observe that wettability switches from underwater oleophilic to underwater superoleophobic. Again, based upon our experimental results, we know that $\gamma_{ow}$ is significantly reduced. If this wetting transition were to simply be due to a reduction in $\gamma_{ow}$, then one would expect $\theta^*$ to decrease, which goes against our experimental finding; therefore, another phenomenon must be at work. From Eq. (1) we see that for $\theta$ to go from $< 90°$ to $> 90°$, then $\Delta\gamma = \gamma_{sw} - \gamma_{so}$ should go from $\Delta\gamma > 0$ to $\Delta\gamma < 0$. The presence of surfactants at the solid-water interface causes a reduction in $\gamma_{sw}$ with respect to $\gamma_{so}$. Previous work on the underwater wetting of oil on low-energy-solids has shown that, for water-soluble surfactants, $\Gamma_{so} < \Gamma_{sw}$ [23]; therefore, one would expect $\gamma_{sw}$ to decrease with respect to $\gamma_{so}$, and for a sufficient decrease, $\Delta\gamma$ should go from $\Delta\gamma > 0$ to $\Delta\gamma < 0$. In addition to this mechanism, which explains the switch in wettability after the addition of surfactant, the small degree of surface roughness (see **Figure S1** for micrographs of surfaces) may also enhance the oleophobicity by allowing small pockets of water to remain entrapped between the oil droplet and the surface resulting in the droplet contacting a composite water-solid interface.[24]

In order to better understand such wetting behavior, it is useful to discuss detergency and its related anti-wetting mechanisms with respect to oil already in contact with a solid. Several mechanisms relevant to the present investigation for oil removal and repellency through the aid of surfactants have been suggested based upon macroscopic experimental observations [7, 8, 25]. The so-called "rolling-up" mechanism (detergency), first proposed by Adam [7], states that the three-phase contact line spontaneously shrinks due to surfactant adsorption at selected system interfaces (oil-water and water-solid), resulting in an increase in contact angle and ultimately oil separation from the surface, and occurs when the following condition is satisfied: $\gamma_{so} + \gamma_{ow} \cos\theta - \gamma_{sw} > 0$ [8]. The ability of the surfactant to infiltrate between the oil and surface and separate the two phases is described by the diffusional mechanism, which has been demonstrated experimentally [8] and with simulations [26, 27] on glass surfaces. (For further information see Supplementary Information, section 'Detergency mechanisms'.) What is relevant to the present study is that, at equilibrium in a surfactant-laden water environment, the hexadecane droplet resides in a non-wetting state with respect to the solid PVDF surface, due to a reduction in $\gamma_{sw}$—switching the wettability of the surface. Interestingly, due to the variety and number of mechanisms available for surfactants to adsorb at a variety of interfaces and inherently change the interfacial properties of the three-phase system, it seems that a wide range of substrates can benefit from this detergency mechanism, and which appears to be supported by the consistently high underwater hexadecane contact angles on all the surfaces that we investigated. All underwater oil contact angles were performed on a substrate that was initially wet with an aqueous solution. If instead, the substrates are initially wet with hexadecane and then exposed to an aqueous solution, the contact angle behavior of hexadecane is different. This is shown by Suppl. Video 1, where the initial contact angle is ~0° and it steadily increases with time. While the contact angle increases, we did not observe a contact angle >150° during the several minutes that the measurement occurred over. In this study, we only used water-soluble surfactants. Were the surfactants to be oil-soluble, one should expect that surfactants would only adsorb at the oil-water and oil-solid interfaces and not the water-solid. Therefore, only $\gamma_{so}$ and $\gamma_{ow}$ will be reduced relative to $\gamma_{sw}$. Applying this to the roll-up condition, $\gamma_{so} + \gamma_{ow} \cos\theta - \gamma_{sw} > 0$, shows that detergency or superoleophobic behavior is less likely.

Ultimately, the presence of surfactant rendered all of our studied surfaces underwater superoleophobic without requiring micro/nanotexturing or chemical treatment. We found that for one to observe underwater superoleophobic behavior for hexadecane, the surfactant concentration should exceed the CMC value (see **Figure S3**). The observation of this phenomenon across the complete range of substrates shows the versatility and facility of this detergency mechanism and opens up the choice, without restriction, of the substrate materials for facile oil separation.

*Oil-Water Separation using Non-functionalized Surfaces*

Our observations of the intrinsic underwater superoleophobic behavior of a range of surface types in a surfactant environment led us to investigate this phenomenon and its implications for the efficacy of oil-water separation through filtering, aiming at surface simplicity, durability and robustness which, in different combinations, are limiting factors of micro- and nanotextured surface technologies (i.e., coating degradation).

The efficacy of a separation medium for oil-water emulsion separation can be determined by its separation efficiency. Separation efficiency for a fixed energy input (e.g., pumping) can be quantified by the overall ability to separate oil and water and the speed at which this can be done. These two quantities end up being intimately related. On the one hand, the ability to completely separate oil and water depends on the smallest oil droplets that can be blocked from passing through the separation medium. Understandably, smaller pore radii, $R$, are able to block smaller oil droplets, ultimately leading to better separation ability for the finest emulsions. The pore radii, however, also exhibit a strong relationship to the pressure drop across the separation medium, where the pressure drop per unit length of the pore, $\Delta P / L$, is proportional to $\propto R^{-4}$, which is an important factor affecting the energy input and separation speed [28]. In short, the smaller the pore size, the slower the allowable flow rate will be for the same pressure drop. Due to mechanical strength limitations of filter media and concerns of oil droplet breakthrough forces, it is advisable to keep the pressure drop across the separation media as low as possible. To this end, we conducted experiments that quantified the optimal pore size in relation to complete separation ability, by quantifying the smallest droplets that can be filtered with a given pore size through the so-called "size-sieving" effect, or separation based upon drop size [6].

Figure 3a illustrates the setup used to conduct these experiments, which consists of (1) an oil (hexadecane)-in-water surfactant (SDS, 0.03 mol L$^{-1}$ in water) stabilized emulsion flowing due to gravity against (2) a simple stainless steel mesh; the interaction between oil droplets and mesh is visualized with (3) an inverted home-built fluorescence bright field laser microscope with a water immersed objective. Note that the surface of the wires in the mesh is not textured; instead, the interaction with the oil is purely regulated by the presence of the surfactant in the mixture. The flow is regulated with (4) a syringe pump. Figure 3b shows a micrograph of one of the as-purchased woven stainless steel meshes that was used in this study, which has repeated square pores. Figure 3c shows a micrograph of a single pore indicating the pore width, $w = 20$ µm; the mesh is false-colored yellow to improve clarity. While the density of hexadecane ($\bar{\rho}$) is less than that of water ($\rho$), and therefore, hexadecane is buoyant, the velocity of water under these circumstances—although flowing in the direction opposite that of the buoyancy force acting on the oil droplet—is sufficient ($u \approx -0.35$ mm s$^{-1}$) to produce a drag force capable of entraining droplets of diameter $d \approx 50$ µm or less (see Supplementary Information, section 'Droplet distribution'). Figure 3d (Suppl. Video 2) and Figure 3e (Suppl. Video 3) show image sequences obtained with fluorescence bright field microscopy, where fluorescent emulsified oil droplets are flowing towards a square pore of width 20 µm—in the case of the large droplet, the droplet is blocked while in the small droplet passes through. In the case where the droplet is blocked (Figure 3d), from Suppl. Video 2 we can see the entire droplet and that it retains a spherical shape during impact with the mesh, indicating that the mesh is completely non-wetting with respect to the droplet even at the microscale. To better understand the minimum value of $d$ that can be blocked by a pore of width $w$, it is instructive to expose the mesh to an emulsion with a distribution of droplets with diameter values in the range of $w$, thus we perform droplet sieving experiments. (More detailed information of the experimental setup and conditions is given in the Methods, section 'Characterization'.) We did not study the sieving behavior of oil-in-water emulsions with no surfactant present. If there is no presence of surfactant, then creating a large number of very small oil droplets, which are difficult to separate, is energetically unfavorable. Therefore, the droplets will migrate to the surface of the water relatively quickly, due to buoyancy forces, where they will pool. At this point, the oil can be skimmed (see Suppl. Video 4).

Figure 4 shows the results of these sieving experiments. For a control experiment, the size distribution of the surfactant-stabilized oil droplets in the flowing emulsion ($u \approx 0.35$ mm s$^{-1}$) was measured without any separation medium (Figure 4, hatched bars). Here, we can see the relative frequency of an oil droplet, $N/N_0$, vs. $d$ ($N_0 = 1,100$). No droplets with $d > 45$ µm are observed, as expected (see Supplementary Information, section 'Droplet distribution'). Figure 4 (gray bars) also shows the distribution of oil droplets ($N/N_0$, vs. $d$) that have passed through a mesh with $w = 20$ µm, and it is clear that there is a strong sieving effect at $d \approx w$. That is, droplets with $d < w$ pass through the mesh while, for $d > w$, the droplets are blocked. It should be noted that during these experiments we observed minor choking of the mesh due to blocked droplets that did not immediately coalesce, however no fouling due to oil wetting the mesh was observed. We expect that, in the actual separation process, blocked droplets will eventually coalesce [29], growing larger and increasing their buoyancy force allowing them to overcome drag and capillary forces, which will limit choking of the pores. This drop coalescence rate, however, remained outside of the scope of this study.

To better understand why $d \approx w$ is a sufficient condition for droplet separation (i.e., minimal droplet deformation), it is instructive to consider the flow behavior of a single emulsified oil droplet. Figure 5a shows the flow conditions around a rigid sphere when $Re = \rho(V_z - u)d/\mu \ll 1$, where $V_z$ is the sphere velocity, $u$ is the uniform flow velocity far from the sphere, and $\mu$ is the viscosity of the surrounding fluid. Here, the sphere experiences a drag force (Stokes's) which is equal to $3\pi d \mu (u - V_z)$ and a buoyance force which is equal to $\pi d^3 (\bar{\rho} - \rho) g / 6$ where $g$ is the acceleration due to gravity where the forces act in opposite directions. Therefore, when the forces are in balance, the sphere has a terminal velocity. In the present work, the sphere is a viscous liquid, and accounting for this, the terminal velocity can be defined as [28] (see Supplementary Information, section 'Droplet distribution')

$$V_z = \frac{1}{3}\frac{(d/2)^2 g}{\nu}\left(\frac{\bar{\rho}}{\rho}-1\right)\frac{\mu+\bar{\mu}}{\mu+\frac{3}{2}\bar{\mu}}+u, \qquad (3)$$

where $\nu = \mu/\rho$ is kinematic viscosity and $\bar{\mu}$ is the viscosity of the sphere. To understand how the surrounding flow affects the sphere velocity, Figure 5b shows a contour plot of $d$ vs. $u$ vs. $V_z$ calculated from Eq. (3) where the surrounding fluid and liquid sphere are water and hexadecane, respectively, and it is clear that there is a range of values for $d$ and $u$ where $V_z < 0$ and is therefore entrained in the flow. To see if the Stokes's estimate of the drag force is valid in Figure 5b when $V_z < 0$ ($|Re| \ll 1$), Figure 5c plots $d$ vs. $u$ vs. $Re$. It is clear that, for the experimental conditions in Figure 3 and Figure 4 ($-u \approx 0.35$ mm s$^{-1}$ and $d \approx 5$ to 40 µm), this condition is satisfied ($Re = 2.1\text{E-5}$ to $0.01$). With $V_z$, we can now estimate the dynamic pressure (driving sphere deformation) and how important it is with respect to Laplace pressure (resisting sphere deformation) for a low-viscosity sphere ($\bar{\mu}(V_z - u)/\gamma_{ow} \ll 1$), which is represented by the Weber number ($We = \bar{\rho}V_z^2 d/\gamma_{ow}$). Large and small values of $We$ indicate that large and small droplet deformation is to be expected, respectively. Figure 5d plots $d$ vs. $u$ vs. $We$ for $V_z < 0$, and for the range of interest for $u$ and $d$ ($-u \approx 0.35$ mm s$^{-1}$ and $d \approx 5$ to 40 µm, ), $We \approx 10^{-8}$ to $10^{-7}$ ( $\bar{\rho} = 0.77$ g cm$^{-3}$ [30], $\gamma_{ow} = 5.3$ mJ m$^{-2}$ (see Figure S2)). Although the surfactant lowers the interfacial tension between the oil and water phases, the small $d$ still causes $We$ to be extremely small at the chosen flow rate and significant droplet deformation is energetically unfavorable. Therefore, the capillary forces are expected to dominate the inertial forces during the separation process, which is confirmed by the sieving experiments in Figure 4. Since the droplet sizes that were observed in the presence of the mesh were almost all smaller than the pore size of the mesh, we can assume that very minimal deformation of the droplet due to inertia to squeeze the drop through the pore took place. We conclude that, to be most efficient in separation, one should choose a value of $w$ that is just slightly smaller than the smallest droplet diameter that is desired to be filtered. After the droplets are blocked by the mesh by the optimal pore size, we expected that oil droplets would coalesce via Ostwald ripening, producing larger droplets (increased $d$) that have a buoyancy force sufficient to overcome the drag force of the entraining water flow and then they would move towards the top of the emulsion, separating the oil from the water phase. (See Figure 5b for a critical value of $d$ necessary to have $V_z > 0$ for a given $\bar{u}$.) Due to the low values of $We$, it is expected that the separation flux could be greatly increased without emulsion droplets passing through the mesh. If we impose the conditions $Re < 1$ and $We \ll 1$, to be consistent with the above model, then from Figure 5c-d we see that the maximum value of $\bar{u}$ is 4.9 mm s$^{-1}$, which corresponds to an emulsion flux rate of 17,640 L m$^{-2}$ hr$^{-1}$.

In order to assess the conditions necessary for droplet break-through to occur when $d > w$, Figure 6 presents a series of experiments where inertia or body forces overcome the resisting surface tension forces to cause an oil droplet to pass through a mesh. Figure 6a-b shows cross-sectional images obtained with X-ray tomography for a series of heavy oil droplets ($\bar{\rho} > \rho$) with a size $d \approx 2.0$ mm (a) before and (b) after exposure to a surfactant-laden environment for meshes with varying $w$ (0.86 mm, 0.38 mm, and 0.10 mm). It is clear that, for the largest value of $w$, the droplet was able to pass through the mesh upon exposure to surfactant, in spite of the fact that the mesh was intrinsically underwater superoleophobic (see Figure S4). In the case of smaller $w$, the deformation due to the addition of surfactant is not sufficient to let the droplet through the mesh. This can be explained by comparing the gravitational pressure ($d(\bar{\rho}-\rho)g$) and Laplace pressure necessary to penetrate through the mesh pore assuming that the mesh is underwater superoleophobic ($4\gamma_{ow}/w$). The ratio of these two is the gravitational Bond number,

$Bo = \dfrac{(\bar{\rho}-\rho)gwd}{4\gamma_{ow}}$. Before the addition of surfactant, $Bo$ = 0.1, 0.04, and 0.01 for $w = 0.86$ mm, 0.38 mm, and 0.10 mm, respectively ($\bar{\rho}-\rho$ = 0.79 g cm$^{-3}$; $d$ = 2.0 mm; $\gamma_{ow}$ = 33 mJ m$^{-2}$). (Refer to Figure S2 for experimental values of $\gamma_{ow}$). After the addition of surfactant, $Bo$ = 0.5, 0.2, and 0.05 for $w = 0.86$ mm, 0.38 mm, and 0.10 mm, respectively, and the droplets become puddle shaped due to the reduction in $\gamma_{ow}$ ($\bar{\rho}-\rho$ = 0.79 g cm$^{-3}$; $d$ = 2.0 mm; $\gamma_{ow}$ = 7.2 mJ m$^{-2}$). For the case where $Bo \geq 0.5$, the mesh was no longer able to support the droplet—in spite of the fact that it was intrinsically underwater superoleophobic (see Figure S4)—and the droplet passed through. Due to limitations in the speed of image acquisition in X-ray tomography, as a complement, we visualized the passing process optically, Figure 6c. Here it is clear that the addition of surfactant causes a reduction in the interfacial tension of the droplet and that the surface tension forces are no longer capable of supporting the droplet on the mesh, resulting in a portion of the droplet passing through the mesh (Suppl. Video 5). While body forces are indeed capable of causing droplets to penetrate pores when $d > w$, inertia can also have a similar effect. Figure 6d plots $We = \bar{\rho}V_z^2 d / \gamma_{ow}$ vs. $d/w$ for FC-770 droplets and jets impacting on meshes in a surfactant laden environment (Triton X-100, 1.3E-3 mol L$^{-1}$), and it is clear that, for sufficiently high values of $We$, oil can penetrate the mesh in spite of the fact that $d > w$. Since high values of $We$ are not readily achievable for droplets under these conditions, we decided to generate liquid jets of oil, of diameter $d$—with the assistance of a pump—and impact them onto the mesh. It is clear that once $We > 1$, $d \approx w$ is no longer a sufficient design rule for blocking oil. Therefore, for controlled filtering and sieving experiments, once should ensure that $We \ll 1$.

*Large-area separation materials*

Up until this point we have created a science base to support the development of a new scalable technology, based on commercially available materials to separate surfactant stabilized oil-in-water emulsions, that is potentially applicable in an industrial setting. The science base is grounded in two areas: thermodynamics (the equilibrium wetting nature of different materials in different environments) and fluidics (the balance between pore size and separation efficiency). Based upon our thermodynamic observations and analysis, we have learned that the detergency effect is applicable to a wide range of materials, allowing a practically unconstrained choice of material for the filter medium. Based upon our fluidics observations and analysis, we have learned the relationship between the pore size of the filter medium to the smallest allowable emulsion droplet size to achieve separation. Using this information we selected non-woven glass/polypropylene syringe filter—with effective pore sizes of 450 nm—as a scalable, large-area, and commercially-available solution to efficiently filter emulsions with minimum droplet diameters in the microscale. To demonstrate this, Figure 7a-d shows gravity-driven emulsion separation with such a non-woven material. In this demonstration, the emulsion flux is ~11,000 L m$^{-2}$ hr$^{-1}$ bar$^{-1}$, which, from the above fluidics considerations, could be increased without suffering droplet deformation and filter penetration (see also Supplementary Information, section 'Separation Efficiency' and Suppl. Video 6).

**Conclusions**

The present study investigated the fundamentals of filtering emulsions that are stabilized in a surfactant-laden water environment. We found that surfactants exhibited a detergency effect beyond a certain concentration (critical micellar concentration), causing oil droplets—at both micro and macroscale—to be in a superoleophobic state with respect to a range of solid materials without the need for surface chemical functionalization or surface micro- or nanotexturing—a wettability design rule. With this new understanding with respect to wettability, we identify the pore size of a woven mesh necessary to filter an emulsion droplet with a prescribed diameter, with the goal being to minimize the pressure drop across the mesh. Due to the small scale of the droplets, we found that capillary forces dominated, and droplet deformation was energetically unfavorable during filtering. Therefore, to filter a droplet with a given diameter, the pore opening should be just slightly smaller—a fluidic design rule. Once the droplets are blocked by the mesh, we expect that coalescence of droplets will occur and, once the droplets are large enough, that they move upwards due to buoyancy to populate the region of oil at the top of the emulsion where they can be skimmed. We estimate that the condition of minimal droplet deformation ($We \ll 1$) and inertial effects ($Re < 1$) should be satisfied until separation (volumetric) fluxes of $\sim 10^4$ L m$^{-2}$ hr$^{-1}$, demonstrating the efficacy of this approach. Due to the detergent-driven non-wetting nature of the meshes, we were able to demonstrate the ability to filter droplets on the basis of size, with the mesh filtering droplets above a selected diameter (emulsion sieving). We then demonstrated complete separation of a surfactant-stabilized emulsion with a commercially available non-woven material (as purchased, no functionalization) with an emulsion flux of $\sim 11,000$ L m$^{-2}$ hr$^{-1}$ bar$^{-1}$ demonstrating the capabilities of this approach.

**Methods**

*Materials*

We used the following chemicals and substrates in our study on wettability under the influence of micellar solutions: poly(vinylidene fluoride) (PVDF, Sigma-Aldrich, $M_w$=180.000 by GPC), poly(methyl methacrylate) (PMMA, EVONIK, high-gloss graphical quality), borosilicate glass D263M (Menzel-Gläser), aluminum (Metall Service Menziken, EN AW-1085 (AI99.9), highly polished), stainless steel (Metal Service Menziken, 304/1.4301, N4 finish), sodium dodecyl sulfate (SDS, Sigma-Aldrich, micellar average molecular weight is 18,000), hexadecane (Sigma-Aldrich, 99%), Triton X-100 solution (Sigma-Aldrich, laboratory grade), Fluorinert FC-770 oil (Sigma-Aldrich), and de-ionized water (Millipore Direct-Q 3 System).

For the separation investigation, we used the following materials: metal meshes (TWP Inc., see Table 1 for the mesh specifications), syringe filter (Acrodisc, glass 1 μm fiber prefilter, 0.45 μm polypropylene filter), fluorescent dye (Radiant Color N.V., RADGLO CFS-0-06 Yellow, oil-soluble, absorption peak wavelength is 440 nm, emittance wavelength is 510nm), oil dye (Sigma-Aldrich, Sudan II, dye content 90%), sodium dodecyl sulfate (per previous specification), Triton X-100 solution (Sigma-Aldrich, laboratory grade), hexadecane (per previous specification), Fluorinert FC-770 oil (Sigma-Aldrich), de-ionized water (per previous specification).

*Preparation*

Prior to contact angle measurements, we cleaned all substrates by sonication in a de-ionized water bath, isopropanol bath, and an acetone bath each for five minute time intervals (excluding the PMMA in acetone, due to its solubility) followed by three minutes of oxygen plasma-ashing at 100W (Diener electronic GmbH Type FEMTO PCCE). The PDVF surface was spun coat onto the clean glass after cleaning and directly used for testing afterward. We prepared a stock solution of SDS (0.03 mol L$^{-1}$) in de-ionized water for facile experimental repeatability with the micellar environment.

Prior to the separation experiments, we cleaned the stainless steel and aluminum meshes by soaking them in a de-ionized water bath, isopropanol bath, and acetone bath. The syringe filters were used as received.

*Characterization*

We conducted contact angles measurements using both a custom home-built setup and a commercially available goniometer (DataPhysics, OCA 35). The home-built setup consisted of a white LED array light source (Thorlabs, LIU104), CMOS camera (Thorlabs, DCC1645C), zoom lens (Thorlabs, MVL7000), liquid bath, syringe pump (SyringePump, NE-1000), syringe (Hamilton 1 mL), plastic needle (Eppendorf, 20 μL microloader), and tubing (IDEX #1533, PEEK, 1/16" outer dia.). For the contact angles of hexadecane shown in **Figure 2**, we conducted at least ten contact angle measurements for each surface in both a pure de-ionized water and a 0.03 mol L$^{-1}$ SDS water environment. Due to the lower density of hexadecane with respect to water, we built an inverted PMMA housing to simplify the underwater contact angle measurements. We initially filled the housing with either de-ionized water or the stock solution of water and SDS and then carefully set the sample in the liquid environment using clean tweezers. We then introduced a hexadecane drop onto the surface through the dispensing needle to obtain an advancing contact angle measurement. We then removed the oil from the surface through the needle to obtain a receding contact angle measurement. We controlled the addition and removal of the oil using a syringe pump. We used image analysis software on recorded videos of this process, done ten different times on different sample locations for each sample. For the contact angles of FC-770 on stainless steel and aluminum we used the same tools in a similar fashion except that the needle was not inverted due to the high density of FC-770 to water. We continually increased the concentration of Triton X-100 in water in order to understand the relationship between surfactant concentration and contact angle. The results are shown in Figure S4. To investigate the effect of droplet size (hydrostatic pressure) on wetting behavior, **Figure S5** shows an image of a relatively large hexadecane droplet in contact with a PMMA substrate in an aqueous surfactant solution (0.03 mol L$^{-1}$). In spite of the large droplet size, and corresponding hydrostatic pressure, no effect on the droplet contact angle was observed. The droplet diameters used for measuring the advancing and receding contact angle values reported in **Figure 2** were kept below the capillary length.

We used the pendant drop method, with a commercially available droplet shape analyzer (DataPhysics, OCA 35) [31], to calculate the interfacial tension of hexadecane in water, SDS, and Triton X-100 micellar solutions with varying concentration. A transparent plastic bath was initially filled with hexadecane and a water droplet with different concentrations of SDS (ranging from 0 to 0.03 mol L$^{-1}$ in water) was introduced through the dispensing needle into the hexadecane environment. We recorded the droplet shape and analysis was done using computer software. Five to ten measurements for each concentration were performed. A pristine plastic housing and needle was used for interfacial tension measurements without surfactant. Between each trial of varying surfactant concentration, the plastic housing was thoroughly cleaned and the needle replaced, with measurements beginning from low-concentrations of surfactant and proceeding to higher values. We used the same methods as above to calculate the interfacial tension of FC-770 in water and Triton X-100 with varying concentrations. In this case, the continuous environment was water and FC-770 was dispensed from a needle. The results are shown in Figure S2.

We tested the separation characteristics of the stainless steel mesh by analyzing droplet distributions of a surfactant stabilized oil-in-water emulsion with and without the presence of the stainless steel mesh. The emulsion had a concentration of ~4 wt.% hexadecane and its flow rate was 382 µL min$^{-1}$ in a pipe of diameter $D = 4.8$ mm; therefore, the mean flow velocity was $u = -0.35$ mm s$^{-1}$ and $Re_D = \rho |u| D / \mu = 1.9$, where $\mu$ is the viscosity of water. The emulsions were generated by mechanically mixing hexadecane and a micellar solution (SDS 0.03 mol L$^{-1}$ in water) at room temperature. In order to perform sieving experiments, we wanted to be able to reproducibly produce a significant number of oil droplets with diameters (*d*) larger than the mesh pore sizes (*d* > *w*). To achieve this, we used the same emulsion mass (10 g), oil concentration (4 wt.% oil), temperature (room temperature), mixing speed (1500 s$^{-1}$), and mixing time (25 min). Furthermore, for sieving experiments, the emulsion was added to the experimental setup immediately after mixing ceased—to minimize the effect of coalescence—and the emulsion flow rate was kept constant (382 µL min$^{-1}$ in a pipe of diameter 4.8 mm). **Figure S6** shows the relative oil droplet diameter frequency for four separate surfactant stabilized oil-in-water emulsions flowing in the experimental setup, and it is clear that there is a significant fraction of oil droplets with *d* > *w*. The same emulsion preparation procedure was used for creating surfactant stabilized oil-in-water emulsions for the dynamic light scattering study, and the oil concentration was 5 wt.% hexadecane.

The experimental set-up consisted of a home-built inverted bright field laser microscope with a water immersed objective (see Figure 3). We used a high-speed camera (Photron FASTCAM SA1.1) to record the separation for later analysis and made a PMMA housing in order to facilitate observations. The housing consisted of a hollow column that was mounted into a recessed hole in a PMMA block which could be fitted with an objective from below in order to observe droplets on the plane immediately below the mesh (see Figure 3a). Initially, we observed the droplet size distribution of an SDS-stabilized oil-in-water emulsion without the use of any mesh. We initially filled the column with the emulsion and used the syringe pump to control the flow rate. We made five separate recordings to analyze the size and count of oil droplets that we observed using a self-developed Matlab image recognition code. This data acted as the control data for size distribution of the emulsion. We then inserted and sealed the 635 stainless steel mesh between the column and the block and we made five further recordings (focusing on the back side of the mesh) of the same emulsion at the same flow rate. We did not observe any effect of the fluorescent dye on the underwater wetting behavior of hexadecane on the mesh. Again, we analyzed the size and count of these droplets using the same Matlab code. The results of these experiments are illustrated in Figure 4.

The separation ability of large-area, commercially available material was investigated using a syringe filter. We filled a syringe with an emulsion consisting of a 5 wt.% of hexadecane in a 0.03 mol L$^{-1}$ SDS water environment. The syringe filter was connected to the syringe and the emulsion was pushed through the filter into an empty glass vile. A sequence of the separation is illustrated in Figure 7. In order to characterize the oil separation ability of this filter, we analyzed emulsion before and after separation with dynamic light scattering (DLS; Malvern DLS Zetasizer Nano ZS). Specifically, we were interested in characterizing droplets with submicron diameters, which we achieved by allowing the emulsion to settle for > 1 hr. **Figure S7**a shows a plot of intensity vs. effective particle diameter for the oil-in-water emulsion before and after filtering (Acrodisc, glass 1 μm fiber prefilter, 0.45 μm polypropylene filter), and it is clear that before filtering, there are particles with submicron diameters, and that after filtering, all particles with $d > 0.45$ µm are separated. To confirm that this signal is attributed to oil droplets, we ran a control experiment where we analyzed the aqueous surfactant solution before and after filtering. **Figure S7**b shows a plot intensity vs. effective particle diameter for the surfactant solution before and after filtering (Acrodisc, glass 1 µm fiber prefilter, 0.45 µm polypropylene filter), and it is clear that in both cases, the signal intensity is well-below that observed for the unfiltered oil-in-water emulsion.

To investigate the three-dimensional wetting configuration of oil droplets lying on the separation materials before and after exposure to a surfactant laden water environment, we performed synchroton X-ray tomography at the TOMCAT beamline at the Swiss Light Source, Paul Scherrer Institute, Villigen, Switzerland. The TOMCAT beamline exploits a 2.9 T magnetic dipole, corresponding to a critical energy of 11.1 keV, generating a closely bundled and highly brilliant X-ray beam. A fixed-exit double crystal multilayer monochromator (DCMM) is set up to obtain a central X-ray photon with polychromatic radiation with a 5% filter, resulting in broadband radiation with a peak energy of 30 keV. The X-ray beam passing through the sample setup is converted by scintillator (LAG:Ce 20 µm) into visible light, which is captured by a 2-4x adjustable magnification objective lens fitted to a sCMOS camera with a pitch of 6.5 µm. The synchrotron X-ray radiography images were recorded at 33 fps, with an exposure time of 7 ms, and at 0.33 mm pixel resolution, providing high temporal and spatial accuracy. The spatial resolution is set by the microscope system of TOMCAT beamline, where the effective camera pixel size is 1.78 µm and the magnification of objective lens is 3.65x. Reconstruction was done using standard absorption contrast.

We obtained the results for the effect of surfactant on the Bond number (Figure 6c) by initially setting a FC-770 oil drop onto the aluminum 20 mesh in a water environment and introducing a 0.06 mol L$^{-1}$ Triton X-100 solution in water at a rate of 0.1 mL min$^{-1}$ with the syringe pump and filming the event with a high-speed camera (Phantom v9.1). We obtained the results for the effect of $We$ on separation limitation (Figure 6d) by filming the impact of FC-770 oil in the form of a droplet and also a jet—in order to achieve high values of $We$—on a variety of meshes with the same high-speed camera. We were able to vary the Weber number by changing the pumping velocity of the syringe pump and therefore the impact velocity and by changing the size of the oil jet or droplet. We were able to vary the ratio of drop diameter to pore opening ($d/w$) by using different meshes and by changing the size of the oil jet or droplet.

**Supporting Information**

Explanation of detergency mechanisms. Calculation of droplet distribution based on flow velocity. Calculation of separation efficiency. Surface characterization. Effect of surfactant on interfacial tension and underwater wetting behavior. Relative frequency of oil droplet diameters in a surfactant stabilized oil-in-water emulsion. Dynamic light scattering characterization of surfactant stabilized oil-in-water emulsions before and after filtering

Videos: Dewetting of oil on PMMA; sieving a surfactant stabilized oil-in-water emulsion: blocking and passing; natural oil separation behavior of oil-water emulsion; oil drop on mesh exposed to surfactant solution; separating a surfactant stabilized oil-in-water emulsion

**Author contributions statement**

D.P., T.M.S., and D.D. conceived the research, designed research and provided scientific guidance in all aspects of the work. T.M.S., C.W., R.S., T.M., S.J., C.S., C.A., H.E., J.L.F., and A.P. conducted the experiments, T.M.S., C.W., and R.S. analyzed the results. D.P., C.W., and T.M.S. wrote the paper draft and all authors participated in manuscript reading, correcting, and commenting.

## Acknowledgements

Partial support of the Swiss National Science Foundation under grant number 162565 and the European Research Council under Advanced Grant 669908 (INTICE) is acknowledged. T.M.S. also acknowledges the ETH Zurich Postdoctoral Fellowship Program and the Marie Curie Actions for People COFUND programme (FEL-14 13-1). Support from PSI through the proposal 20140203 is acknowledged. T.M.S. and C.W. acknowledge Anastasia Spyrogianni and Dr. Robert Büchel for their assistance with the DLS measurements.

## Additional information

**Competing financial interests:** The authors declare no competing financial interests.

**Figures**

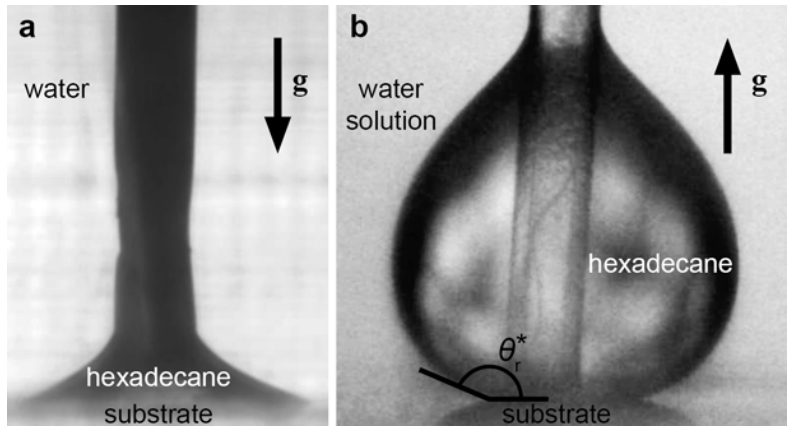

**Figure 1. Surfactants (detergents) can switch the wettability of a surface.** Image sequences showing an underwater receding contact angle measurement (withdrawal method) of hexadecane on a poly(vinylidene fluoride) coated substrate **a**, without, and **b**, with surfactant in the water (0.03 mol L$^{-1}$ sodium dodecyl sulfate in water). Arrows indicate the direction of acceleration due to gravity (**g**). Due to the buoyancy of hexadecane in water, the image in **b** is inverted to enhance clarity.

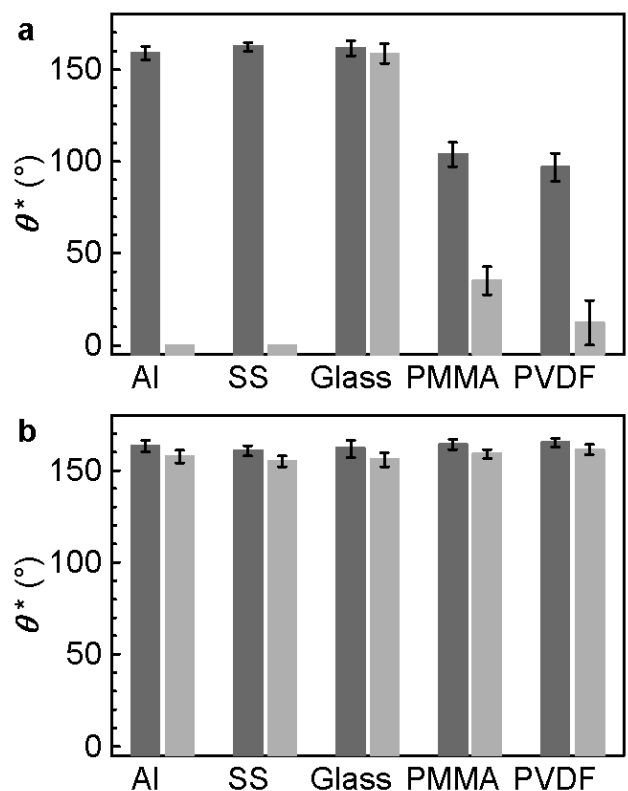

**Figure 2 Detergents alter the underwater wetting behavior of oils to a range of materials.** Underwater advancing (dark gray bars, $\theta_a^*$) and receding (light gray bars, $\theta_r^*$) contact angle measurements with hexadecane on a range of substrate materials **a**, without and **b**, with the presence of a surfactant (sodium dodecyl sulfate, 0.03 mol L$^{-1}$). Acronyms: Al, aluminum; SS, stainless steel; PMMA, poly(methyl methacrylate); PVDF, poly(vinylidenefluoride).

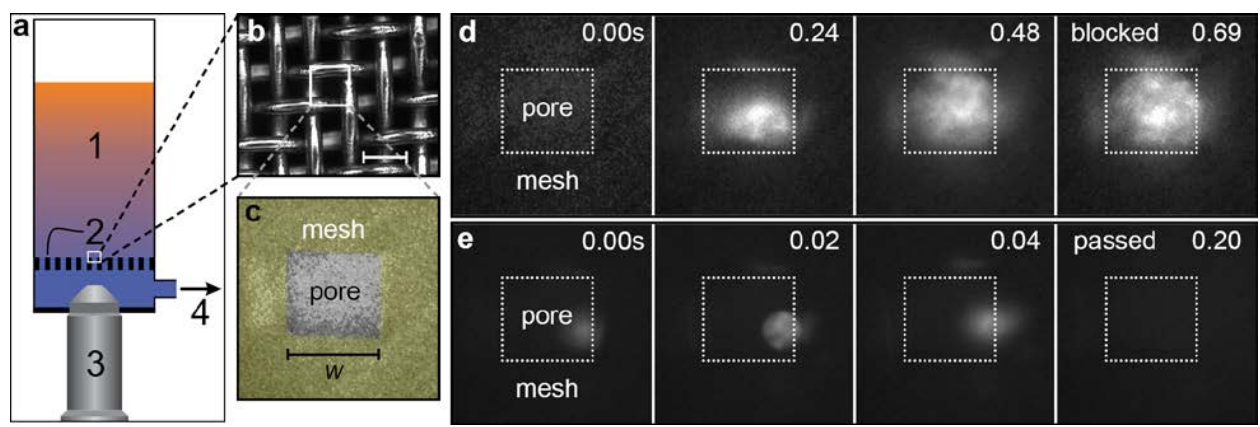

**Figure 3 Demonstrating emulsion sieving**. **a**, Schematic showing the experimental setup used to investigate emulsion sieving: (1) Surfactant stabilized oil-in-water emulsion (SDS 0.03 mol L$^{-1}$); (2) metallic woven mesh; (3) inverted water-immersed objective; (4) sieved emulsion exit flow controlled with a syringe pump. Micrographs of the mesh used at **b**, low- and **c**, high-magnification; false coloring is used to highlight the mesh and pore regions. The fiber spacing is $w = 20.3$μm and the diameter is 20 μm. High-speed fluorescence imaging of luminescent oil droplets (dyed) **d**, being blocked by and **e**, passing through the mesh (see also Suppl. Video 2 and Suppl. Video 3). Dotted line indicates the approximate region of the mesh pore. Scale bar: **b**, 50 μm. See Figure 5 for a description of the surrounding flow speed.

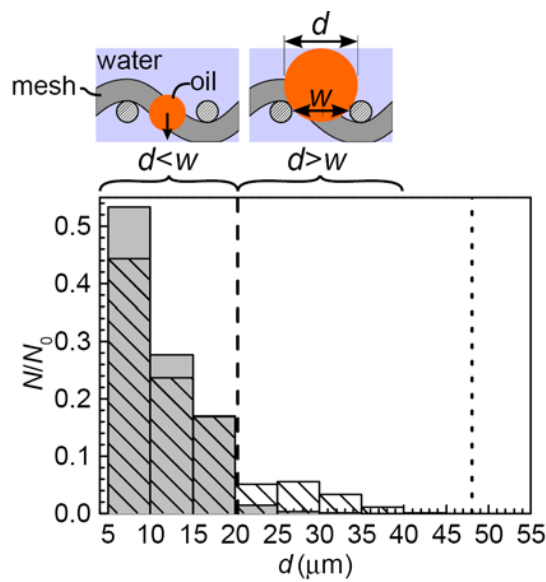

**Figure 4 Standard metallic meshes are capable of sieving emulsions without fouling.** Plot of the relative frequency of a droplet in a surfactant stabilized oil-in-water emulsion ($N/N_0$) vs. oil droplet diameter ($d$) with (filled gray bar, total droplet count: $N_0 = 1,100$) and without the presence of a mesh (hatched bar, $N_0 = 1,180$). Here, the emulsion flows in the direction opposite to that of the force due to buoyancy acting on the oil droplets ($\bar{\rho} < \rho$, where $\bar{\rho}$ and $\rho$ are the density of oil and water, respectively). The two vertical lines represent the wire spacing of the mesh (– – –; $w = 20.3\,\mu\text{m}$) and the value of $d$ where the oil droplet velocity is zero, $V_z = 0$ (see Figure 5 and Supplementary Information, section 'Droplet distribution'). Also shown are the regions where the droplets are smaller ($d < w$) and larger ($d > w$) than the wire spacing and the corresponding behavior of the droplet, i.e., passing ($d < w$) and blocking ($d > w$). The emulsion had a concentration of ~4 wt.% hexadecane and its flow rate was 382 µL min$^{-1}$ in a pipe of diameter $D = 4.8$ mm; therefore, the mean flow velocity was $u = -0.35$ mm s$^{-1}$ and $Re_D = \rho|u|D/\mu = 1.9$, where $\mu$ is the viscosity of water. Surfactant concentration: SDS 0.03 mol L$^{-1}$.

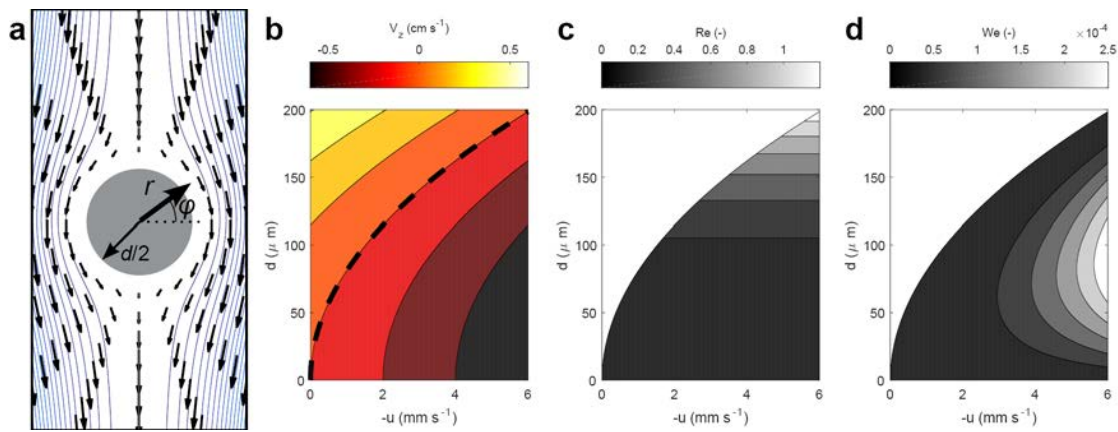

**Figure 5 Effect of liquid flow velocity on the critical diameter of an oil droplet that is entrained**. **a**, A rigid sphere of diameter $d$ in a fluid with a uniform velocity, $u$. Also shown is the resulting velocity vectors, $\mathbf{u}/u$, and isolines for the Stokes's solution for flow around the sphere, which comes from the iso-stream function, $\bar{\psi} = \bar{r}^2 \sin^2\phi \left( \frac{1}{2} - \frac{3}{4\bar{r}} + \frac{1}{4\bar{r}^3} \right)$, where $\bar{r} = 2r/d$, $Re_{\text{flow}} = \rho u d / \mu \ll 1$, $\frac{u_r}{u} = \frac{1}{\bar{r}^2 \sin\phi} \frac{\partial \bar{\psi}}{\partial \phi}$, and $\frac{u_\phi}{u} = -\frac{1}{\bar{r} \sin\phi} \frac{\partial \bar{\psi}}{\partial \bar{r}}$ (see Ref. [32] for solution). When $u < 0$ and the sphere is less dense than the surrounding fluid, the resulting drag force acts against the buoyance force. Uniform flow velocity, $u$, vs. hexadecane droplet diameter, $d$, vs. **b**, hexadecane droplet velocity, $V_z$ **c**, Reynolds number, $Re = \rho(V_z - u)d/\bar{\mu}$, of the flow around the droplet, and **d**, Weber number, $We = \bar{\rho} V_z^2 d / \gamma_{\text{ow}}$, of the droplet. Dashed line in **b,** indicates $V_z = 0$. **c**, and **d**, are only shown for $V_z \leq 0$ (i.e., droplets entrained in the surrounding flow). The flow speed of $-u = 6$ mm s$^{-1}$ corresponds to an emulsion flux rate of 21,600 L m$^{-2}$ hr$^{-1}$.

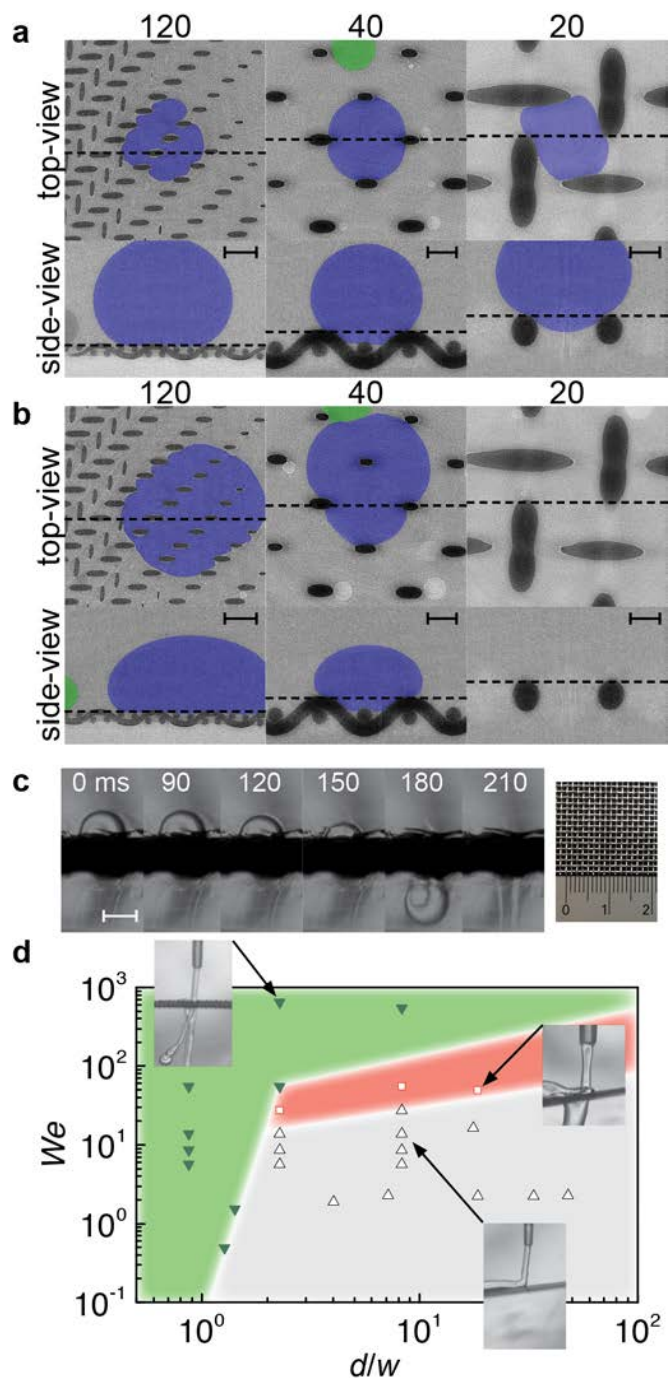

**Figure 6 The effect of surfactant and substrate on the wetting behavior of oil (FC-770) on metallic meshes**. Top-view and side-view cross-sectional images of a dense oil droplet resting on a series of aluminum meshes (120, 40, and 20 pores per inch) in a water environment **a**, before and **b**, after the addition of a water-surfactant mixture (Triton X-100 in water); the mixture was added continuously until significant deformation of the droplet occurred. The wire diameters are 0.41 mm (20 mesh), 0.25 mm (40 mesh), and 0.10 mm (120 mesh) and the wire spacings are 0.86 mm (20 mesh), 0.38 mm (40 mesh), and 0.10 mm (120 mesh). Oil droplets are false colored blue and green for clarity. **c**, Image sequence illustrating how the addition of surfactant (Triton X-100) to a water bath can cause a dense oil droplet to pass through a pore in an aluminum 20 mesh. (See

Methods section for a description of the concentration and rate of introduction of surfactant addition and Suppl. Video 5.) A picture of the mesh is also included for better understanding. **d,** Weber number ($We = \bar{\rho} V_z^2 d / \gamma_{ow}$) vs. the ratio of jet or droplet diameter ($d$) and wire spacing ($w$); the density of FC-770 is $\bar{\rho} = 1.79$ g cm$^{-3}$ and $\gamma_{ow}$ = 7.3 mJ m$^{-2}$ with and 33 mJ m$^{-2}$ without surfactant (0.0013 mol L$^{-1}$ Triton X-100 in water). Upside-down triangles represent penetration through the mesh. Squares represent results in the transition regime. Triangles represent rebound and mobility of the oil. Each of these three regimes is split up by color; gray predicts mobility, red predicts transition, and green predicts penetration. The images are representative examples of a specific experiment in each regime. Scale bar: **a-b**, 0.5 mm, **c**, 1 mm.

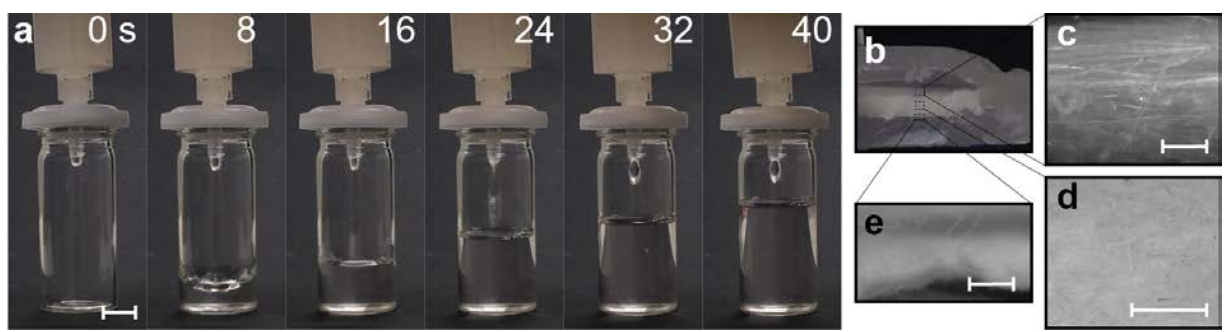

**Figure 7 Demonstrating complete separation of oil from water with commercially available materials.** Surfactant stabilized (SDS) oil (hexadecane)-in-water emulsion (0.03 mol L$^{-1}$ SDS in water, ~5 wt.% hexadecane) being filtered with a commercially available filter. Here, the flow-rate of the water and emulsion is controlled with a syringe pump and the separation flux is ~11,000 L m$^{-2}$ hr$^{-1}$ bar$^{-1}$. **a,** Time series of separation. **b,** Cross-section of filter showing three individual layers. **c-d,** prefilter layers (glass) and **e,** polypropylene. Scale bar: **a**, **c**, **d**, **e**, 100 μm. See Suppl. Video 6.

**Tables**

| Material | wires per inch | wire diameter (mm) | opening ($w$) (mm) |
|---|---|---|---|
| Aluminum | 20 | 0.41 | 0.86 |
| Aluminum | 40 | 0.25 | 0.38 |
| Aluminum | 120 | 0.10 | 0.10 |
| 316/1.4401 stainless steel | 40 | 0.17 | 0.47 |
| 316/1.4401 stainless steel | 90 | 0.09 | 0.19 |
| 316/1.4401 stainless steel | 230 | 0.04 | 0.07 |
| 316/1.4401 stainless steel | 635 | 0.02 | 0.02 |

**Table 1 Specifications of the meshes used in the separation investigation.** Note that all wires have smooth surfaces.